\documentclass[12pt]{article}

\usepackage{graphicx}
\begin{document}

\begin{center}

\centerline{\Large\bf Density Dependence of Nucleon Bag Constant,}

\centerline{\Large\bf Radius and Mass in an Effective Field }

\centerline{\Large\bf Theory Model of QCD\footnote{Work supported by the
National Natural Science Foundation of China and the Foundation for
University Key Teacher by the Ministry of Education}}

\vspace*{1cm}

Yu-xin Liu$^{a,b,c,d,e)}$, Dong-feng Gao$^{a)}$ and Hua Guo$^{b)}$

$^a$ Department of Physics, Peking University, Beijing 100871, China

$^b$ The Key Laboratory of Heavy Ion Physics, Ministry of Education,
China, Peking University, Beijing 100871, China

$^c$ Institute of Theoretical Physics, Academia Sinica, Beijing 100080,
China

$^d$ Center of Theoretical Nuclear Physics, National Laboratory of
Heavy Ion Accelerator, Lanzhou 730000, China

$^e$ CCAST(World Lab.), P. O. Box 8730, Beijing 100080, China


\end{center}

\begin{abstract}
With the global color symmetry model (GCM) being extended to finite
chemical potential, the density dependence of the bag constant,
the total energy and the radius of a nucleon, as well as the
quark condensate in nuclear matter are investigated.
A maximal nuclear matter density for the existence of the bag with
three quarks confined within is obtained.
The calculated results indicate that, before the maximal density is
reached, the bag constant, the total energy of a nucleon and the quark
condensate decrease gradually, and the radius of a nucleon increases,
with the increasing of the nuclear matter density. Nevertheless no
sudden change emerges. As the maximal nuclear matter density is reached,
a phase transition from nucleons to quarks takes place and the chiral
symmetry is restored.

\end{abstract}

\bigskip



\newpage

\parindent=20pt

\section{Introduction}

It is well known that the nucleonic and mesonic degrees of freedom
play essential roles in describing properties of nuclear matter and
finite nuclei. Meanwhile, nucleons bound in nuclear medium alter their
properties from those in free space. Although how much those properties
change is of fundamental interest in nuclear physics, it is still not
clear now. Then, in the relativistic mean field theory (RMF) based on
the $\sigma -\omega$ model, which is currently regarded as a standard
model of nuclear physics, the effective nucleon mass in nuclear medium
is determined in self-consistent iteration on the fields of nucleons
and mesons (see, for example, Ref.\cite{SW97}). However, this model
is valid only at the hadronic level, at which a nucleon is described
as a point-like particle. Because of the importance of the substructure
of nucleon observed in deep inelastic scattering experiments and
predicted by quantum chromodynamics (QCD), it is imperative to employ
the quark and gluon degrees of freedom to describe nuclear phenomena.
Unfortunately, it is now very difficult to make quantitative predictions
with QCD at low and intermediate energy region.
Therefore various phenomenological models based on the QCD assumptions,
such as the bag models\cite{Jaf745,FL778}, quark-meson coupling (QMC)
model\cite{Gui88}, and so on, have been developed. In bag models,
the fundamental quantities to characterize a nucleon is the bag constant,
the radius and the energy of the bag. In the QMC model, nucleons are
regarded as non-overlapping bags interacting through the exchange of
mesons. The size and mass of a nucleon are represented by the radius
and energy of the bag, respectively, too. In order to take the
modification of the volume energy on the mass of a nucleon into account,
the bag constant is also exploited in the QMC model.
Originally, the bag constant held fixed as its free-space value in bag
models\cite{Gui88}, then a density dependent bag constant is introduced
by a phenomenological dependence on the medium
density\cite{LTh98,JJ967,MJ978,Guo99,Su990}. However, a sophisticated
QCD foundation of the density dependence of the bag constant, the radius
and the mass of a nucleon is still lacking.

Recently, an effective field theory of QCD, namely the global
color symmetry model (GCM), has been
developed\cite{CR858,FT927,Tan97,LLZZ98}, and shown to be quite
successful in describing hadron properties in free space(i.e., at
temperature $T=0$, chemical potential $\mu =0$). With the
Dyson-Schwinger equation approach of QCD being extended to finite
temperature, the deconfinement and chiral symmetry restoration,
the $\pi$ and $\rho$ meson properties, and a part of the baryon
properties at finite-T and finite-$\mu$ have been investigated
\cite{BR96,BPRS97,BRS98,MRS98,BGP99,HS99}. With the
nontopological-soliton ansatz\cite{FL778}, some parameters related
to the nucleon structure, such as the bag constant, had also been
introduced in the GCM at $T=0$ and $\mu =0$ in a natural
way\cite{CR858}. It seems that a QCD foundation for the bag models
is proposed at the level of the GCM model. With the global color
symmetry model being extended to finite chemical potential $\mu$,
the dependence of the nucleon properties, such as the bag
constant, the total energy and the radius of a nucleon on the
nuclear medium density will be studied in this paper.

The paper is organized as follows. In Section 2 we describe the formalism
of the GCM model at finite chemical potential $\mu$ and the model of the
relation between the chemical potential of quarks and the baryon density
in nuclear matter. In Section 3 we represent the calculation and the
obtained results of the bag constant, the bag radius and the bag energy,
as well as the quark condensate as functions of the nuclear matter density.
It contains also discussions on the results. In Section 4, a brief summary
and some remarks are given.

\section{Formalism}

It has been known that the action of the GCM in free space (i.e., at
chemical potential $\mu = 0$) for the zero mass quark is given\cite{CR858}
in the Euclidean space as
$$
S=\int d^{4}xd^{4}y\overline{q}(x)\left[\gamma \cdot \partial \delta(x-y)
+ \Lambda^{\theta}B^{\theta}(x,y)\right]q(y) +\int d^{4}xd^{4}y
\frac{B^{\theta}(x,y)B^{\theta}(y,x)}{2g^2D(x-y)} \, , $$
where $B^{\theta}(x,y)$ is a bilocal field and $\Lambda^{\theta}$ are
matrices of the Fierz transformation among the spin, color, and flavor
spaces of the quarks. Extending it to finite chemical potential $\mu$,
we have the action of the GCM at finite chemical potential as
$$
S=\int d^{4}xd^{4}y\overline{q}(x)\left[\gamma \cdot \partial \delta(x-y)-
\mu\gamma_{4} + \Lambda^{\theta}B^{\theta}(x,y)\right]q(y)
+\int d^{4}xd^{4}y\frac{B^{\theta}(x,y)B^{\theta}(y,x)}{2g^2D(x-y)} \, , $$
from which the generating functional is defined as
$$Z[\overline{\eta},{\eta}]=\int D\overline{q} Dq DB^{\theta}
e^{[-S+\overline{\eta}q+\overline{q}\eta]} \, . $$
After integrating the quark fields, we obtain
\begin{equation}
S=-\mbox{Tr}{\log\left[\gamma\cdot \partial\delta(x-y)-\mu\gamma_{4}+
\Lambda^{\theta}B^{\theta}\right]}
+\int d^{4}xd^{4}y \frac{B^{\theta}(x,y)B^{\theta}(y,x)}{2g^2D(x-y)} \, .
\end{equation}
Generally, the bilocal field $B^{\theta}(x,y)$ can be written
as\cite{LLZZ98}
\begin{equation}
B^{\theta}(x,y) = B^{\theta}_0(x,y) + \sum_{i}\frac{B^{\theta}_0(x,y)}
{f_{i}} \phi^{\theta}_{i}(\frac{x+y}{2})  \, ,
\end{equation}
where $B^{\theta}_{0}(x,y) = B^{\theta}_0(x-y)$ is the vacuum configuration
of the bilocal field. In the lowest order approximation with only the
Goldstone boson being taken into account, the $\phi^{\theta}_{i}$ includes
$\sigma$ and $\pi$ mesons, and $f_i$ ($i = \sigma, \pi$) stands for the
decay constant of $\sigma$, $\pi$ mesons, respectively.
The vacuum configuration can be determined by the saddle-point condition
$\frac{\partial S}{\partial B^{\theta}_0}=0$, and a equation of the
translation invariant quark self-energy $\Sigma (q,\mu )$ is obtained as
\begin{equation}
\Sigma(p,\mu)=\int \frac{d^{4}p}{(2\pi)^4}g^2D(p-q)\frac{t^{a}}{2}
\gamma_{\nu} \frac{1}{i\gamma \cdot q-\mu\gamma_{4}+\Sigma(q,\mu)}
\gamma_{\nu} \frac{t^{a}}{2}\, .
\end{equation}
With $\tilde{q}_{\mu}=(\vec{q},(q_4+i \mu))$ being introduced, Eq.~(3) can
be rewritten as
$$\Sigma(p,\mu)=\int \frac{d^{4}q}{(2\pi)^4}g^2D(x-y)\frac{t^{a}}{2}
\gamma_{\nu}\frac{1}{i\gamma \cdot \tilde{q}+\Sigma(q,\mu)}\gamma_{\nu}
\frac{t^{a}}{2}\, . $$
Taking the conventional decomposition for the quark self-energy
\begin{equation}
\Sigma(p,\mu)=i[A(\tilde{p})-1]\gamma \cdot \tilde{p}+VB(\tilde{p})\, ,
\end{equation}
where $V=\sigma+i\vec{\pi} \cdot \vec{\tau}\gamma_5$ with restriction
$\sigma ^2 + \vec{\pi} ^2 = 1 $, one can fix the self-energy
$\Sigma(p, \mu)$ by solving the rainbow Dyson-Schwinger equations
\begin{equation}
\left[A(\tilde{p})-1\right]\tilde{p}^2 = \frac{8}{3}\int
\frac{d^{4}q}{(2\pi)^4} g^2 D(p-q)\frac{A(\tilde{q})\tilde{q}\cdot
\tilde{p}} {A^2(\tilde{q}) \tilde{q}^2+B^2(\tilde{q})}  \, ,
\end{equation}
\begin{equation}
\qquad \qquad B(\tilde{p})=\frac{16}{3}\int\frac{d^{4}q}{(2\pi)^4}g^2
D(p-q) \frac{B(\tilde{q})}{A^2(\tilde{q})\tilde{q}^2+B^2(\tilde{q})}  \, .
\end{equation}
Basing on the solution of the Dyson-Schwinger equations, one can determine
the bilocal field, and fix further the GCM action.

With a nontopological-soliton ansatz, the action of the bag models at
finite chemical potential $\mu$ in terms of the GCM can be given by
extending the formalism proposed in Ref.\cite{CR858} as
\begin{equation}
S_B=\sum_{i=1}^{3}\overline{q}_{i}[i\gamma \cdot \partial+\mu\gamma_0-
\alpha (\sigma(x)-i\vec{\pi}(x) \cdot \vec{\tau}\gamma_5) ] q_{i}
+\hat{S}(\sigma,\pi,\mu) \, ,
\end{equation}
where $\hat{S}(\sigma,\pi,\mu)$ includes only $\sigma$ and $\pi$ mesons
and reads
\begin{equation}
\hat{S}(\sigma,\pi,\mu)=\int[\frac{f_{\sigma}^2}{2}(\partial_{\mu}\sigma)^2
+\frac{f_{\pi}^2}{2}(\partial_{\mu} \vec{\pi})^2-V(\sigma,\pi)]d^{4}z+\cdots
\end{equation}
with
$$\displaylines{\hspace*{3mm}
V(\sigma,\pi)=-\frac{12\pi^2}{(2\pi)^4} \int_{0}^{\infty}s'ds'\left\{
\log\left[\frac{A^{2}(s') s'+(\sigma^2+\vec{\pi}^2)B^{2}(s')}{A^{2}(s')s'
+B^{2}(s')} \right] -\frac{(\sigma^2+\vec{\pi}^2-1)B^{2}(s')}{A^{2}(s')s'
+B^{2}(s')} \right\}\, .  \cr } $$
Analyzing the stationary property of the bag and differentiating Eq.~(7),
one has equations for the quarks and mesons
\begin{equation}
\left[i\gamma \cdot \partial +\mu\gamma_0 -\alpha(\sigma(x)-i \vec{\pi}(x)
\cdot \vec{\tau}\gamma_5)\right] q_{i} = 0 \, ,
\end{equation}
\begin{equation}
\frac{\partial S_B}{\partial \sigma(x)} = 0 \,
\end{equation}
\begin{equation}
\frac{\partial S_B}{\partial \pi(x)} = 0\, .
\end{equation}
The quark field and $\sigma$, $\pi $ meson fields in symmetric nuclear
matter can be determined by solving the Eqs.~(9-11) self-consistently.
As a consequence, the corresponding energies can be obtained.
From the restriction on the quark self-energy (Eq.~(4)), it is apparent
that the meson fields corresponding to the vacuum configuration can be
simply taken as $\sigma =1$, $\pi = 0$ under mean-field approximation.
In light of the nontopological-soliton ansatz\cite{FL778,CR858}, one
can take the meson fields as $\sigma=0 $ and $\pi=0$ inside a bag
(i.e., in a nucleon). For the quarks in a bag, Eq.~(9) can thus be
rewritten in the mean-field approximation as
\begin{equation}
\left[ i\gamma \cdot \partial+\mu\gamma_0 \right] q_i(x) = 0
\end{equation}
The lowest total energy of a single quark with respect to the radius $R$
of the bag is given as
\begin{equation}
\epsilon_{i}(R)=\frac{\omega_0}{R} ,
\end{equation}
where $\omega_0=2.04 .$  This is consistent with the results obtained
in Refs\cite{Jaf745,LTh98,JJ967}. And the bag constant ${\cal{B}}$ is
obtained as
\begin{equation}
{\cal{B}}=\frac{12\pi^2}{(2\pi)^4} \int_{\mu^2}^{\infty}s'ds'
\left\{\log\left[ \frac{A^{2}(s')s'+B^{2}(s')}{A^{2}(s')s'}\right]
-\frac{B^{2}(s')}{A^{2}(s')s'+B^{2}(s')}  \right\} \, ,
\end{equation}
with $s^{\prime} = \tilde{p}^2$.
With the correction from the motion of center-of-mass, the zero-point
effect and the color-electronic and color-magnetic interactions being
taken into account, the total energy of a bag is given as
\begin{equation}
E=3\epsilon(R)+\frac{4}{3}\pi R^{3}{\cal{B}}-\frac{Z_0}{R}
 = \frac{3\omega_0-Z_0}{R}+\frac{4}{3}\pi R^{3}{\cal{B}} \, ,
\end{equation}
where $ Z_0/R$ denotes the corrections of the motion of center-of-mass,
zero-point energy and other effects.

Just as the same as that in Ref.\cite{CR858}, the bag is identified as
a nucleon in the present work. It satisfies then the equilibrium condition
$$ \frac{dE(R)}{dR} = 0 \, . $$
From this condition, we get
\begin{equation}
R=\left(\frac{a}{4\pi {\cal{B}}}\right)^{1/4}
\end{equation}
where $ a=3\omega_0-Z_0 . $
As a consequence, Eq.~(15) can be rewritten as
\begin{equation}
E=\frac{4a}{3} \left( \frac{4\pi {\cal{B}}}{a} \right)^{1/4} \, .
\end{equation}

It is apparent that, with the solutions of Dyson-Schwinger equations
(Eqs.~(5) and (6)) being taken as the input for Eqs.~(14), (16) and (17),
the properties of nucleons (i.e., bags) in nuclear matter can be obtained.
Because the quark condensate has commonly been taken as the order parameter
to characterize the phase transition for chiral symmetry restoration
\cite{MK978,ZLGCZ99}, we evaluate also the variation of the quark condensate
against the density of nuclear matter
\begin{equation}
<\overline{q}q>=<:\overline{q}(0)q(0):>=Tr{S(x,x)}
 = - \frac{12}{(2 \pi)^4} \int _{\mu^2} ^{\infty} s^{\prime}
 d s^{\prime} \frac{B(s^{\prime})} { A^2(s^{\prime}) s^{\prime}
 + B^2 (s^{\prime}) } \, .
\end{equation}

In the practical calculation, since the knowledge about the exact
behavior of $ g^2$ and $D(p-q)$ in low energy region is still lacking,
one has to take some approximations or phenomenological form to solve
the Dyson-Schwinger equations. For simplicity, we adopt the infrared
dominative form\cite{MN83,CR858}
\begin{equation}
g^2D(p-q)=\frac{3}{16}\eta^2 \delta(p-q) ,
\end{equation}
where $\eta$ is a energy-scale parameter and can be fixed by experiment
data of mesons. Although this form does not include the contribution from
the ultraviolet energy region, it maintains the main property of QCD in
the low energy region. With Eqs.~(5), (6) and (19), one has
$$\displaylines{\hspace*{1cm}
A(\tilde{p})=2, \qquad \qquad \qquad \qquad \qquad
B(\tilde{p})=(\eta^2-4\tilde{p}^2)^{1/2}, \qquad
\mbox{for} \quad \tilde{p}^2< \frac{\eta^2}{4}, \hfill{(20a)}
\cr \hspace*{1cm}
A(\tilde{p})=\frac{1}{2}\left[1+\left( 1+ \frac{2\eta^2}{\tilde{p}^2}
\right)^{1/2} \right], \ \ B(\tilde{p})=0, \qquad \qquad \qquad \quad
\mbox{for} \quad \tilde{p}^2 > \frac{\eta^2}{4}.
\hfill{(20b)} \cr }  $$

In order to investigate the dependence of nucleon properties on the
nuclear matter density $\rho$ explicitly, we must transfer the above
obtained $\mu$-dependence to that of the $\rho$-dependence.
Because of the fermionic properties of quarks, the bags can be
approximately regarded as a Fermi-Dirac systems\cite{CP75} with the
fermion number density
\setcounter{equation}{20}
\begin{equation}
n=g \int_{0}^{k_F} \frac{d^{3}\vec{k}}{(2\pi)^3 } \, ,
\end{equation}
where g is the degenerate factor, and $ k_F $ is the Fermi momentum.
For the quarks in a nucleon, g is 12, and
$$ k_F=(\mu^{2}-m_{q}^2)^{1/2} , $$
where $m_q$ is the current quark mass. In case of zero current quark mass,
one has $k_F = \mu$. In the lowest order approximation, we get the quark
number density as
\begin{equation}
n_{q}=\frac{2}{\pi^{2}} \mu^3 \, .
\end{equation}

Considering the fact that the baryon number for a quark is 1/3, one can
take the relation between the nuclear matter density and the chemical
potential as
\begin{equation}
\rho_{_B} = \frac{n_q}{3}=\frac{2}{3\pi^2}\mu^3 \, .
\end{equation}

Combining Eqs.~(14), (16-18), (20) and (23), we can obtain the dependence
of the bag constant ${\cal{B}}$, the total energy $E$ and the radius $R$
of a nucleon (i.e., those of a bag) and the quark condensate on the
nuclear matter density $ \rho_{_B}$.

\section{Calculation and Results}

By calibrating the nucleon mass $M_0=939$~MeV as the total energy
of a bag and radius $R_0 =0.8$~fm in free space (i.e., $\mu =0$,
$\rho =0$), we get the energy-scale $\eta=1.220$~GeV, $Z_0=3.303$.
Such best fitted energy-scale $\eta$ fits well the value
$1.37$~GeV, which was fixed by a good description of $\pi$ and
$\rho$ meson masses\cite{BRS98,MRS98}, and is much more close to
the Bjorken-scale $1.0$~GeV (see Ref.\cite{CP75} and the
references therein). The obtained $ Z_0$ is larger than the
originally fitted value $1.84$\cite{HK78}. However what we refer
to here includes all the effects but not only the zero-point
energy. Meanwhile other investigations (see for example
Ref.\cite{MK01}) have shown that the zero-point energy parameter
can be larger than 1.84, even though the other effects are taken
into account separately. With the above parameters $\eta$, $Z_0$
and Eqs.(17) and (18), we get at first the bag constant, and the
quark condensate in free space as $ {\cal{B}}_0
=(172~\mbox{MeV})^4$, and $
<\overline{q}q>_0=-(132~\mbox{MeV})^3$. It is evident that these
obtained values $B_0$ and $<\bar{q}q>_0 $ are quite close to the
results given in Ref.\cite{JJ967} and Ref.\cite{MK978},
respectively.

By varying the chemical potential $\mu$, we obtain the relation between
the nuclear matter density and the chemical potential of the quarks,
which exhibits the same monotonousness apparently.
Furthermore, we get the variation behavior of the ratio of bag constant,
the nucleon radius, the total energy of the bag and the quark condensate
in nuclear matter to the corresponding value in free space against the
nuclear matter density. The results are illustrated in Figs.~1-4,
respectively.

\begin{figure}
\begin{center}
\resizebox{10cm}{!}{\includegraphics{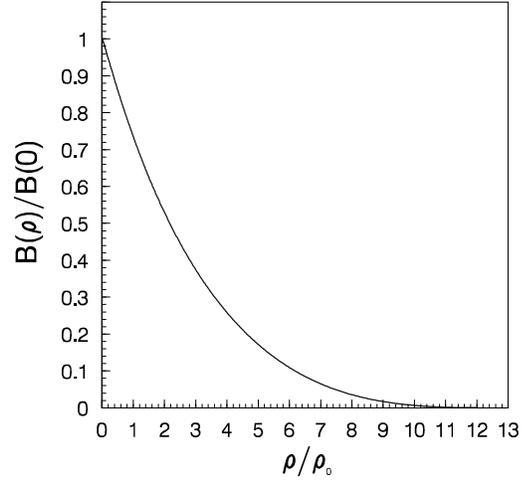}}

\caption{Calculated ratio between the bag constant in nuclear
matter and that in free space as a function of the nuclear matter
density} \label{fig:Fig.1}
\end{center}
\end{figure}

\begin{figure}
\begin{center}
\resizebox{10cm}{!}{\includegraphics{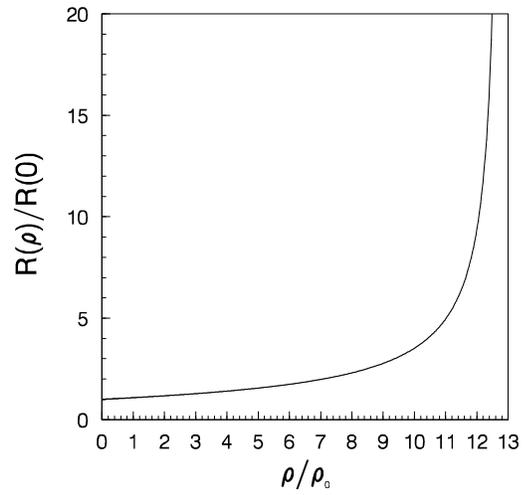}}
\caption{Calculated ratio between the radius of nucleon in nuclear
matter and that in free space as a function of the nuclear matter
density}
\label{fig:Fig.2}
\end{center}
\end{figure}

\begin{figure}
\begin{center}
\resizebox{10cm}{!}{\includegraphics{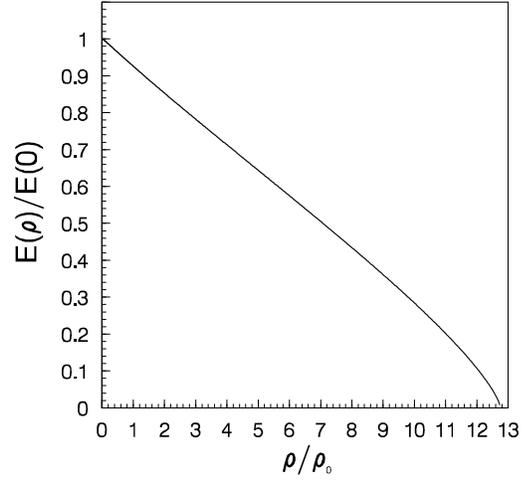}}
\caption{Calculated ratio between the total energy of a bag in
nuclear matter and that in free space as a function of the nuclear
matter density}
\label{fig:Fig.3}
\end{center}
\end{figure}

\begin{figure}
\begin{center}
\resizebox{10cm}{!}{\includegraphics{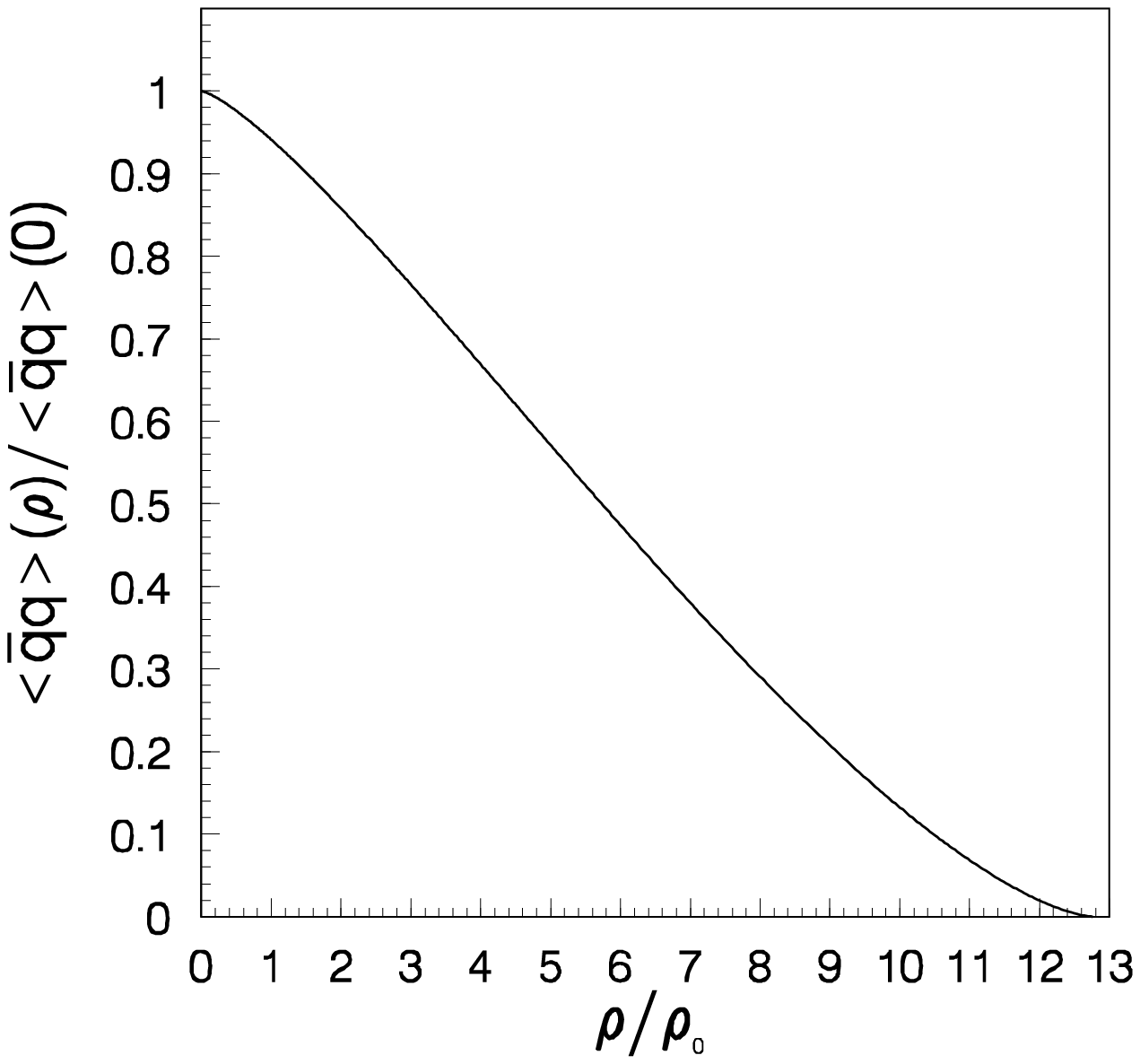}}
\caption{Calculated ratio between the local quark condensate in
nuclear matter and that in free space as a function of the nuclear
matter density} \label{fig:Fig.4}
\end{center}
\end{figure}

Looking over the figures, one may easily realize that, as the density
of nuclear matter increases, the bag constant, the total energy of
the bag and the quark condensate decrease monotonously.
Meanwhile, the radius of a nucleon increases. When the nuclear matter
density reaches the value larger than 12 times the normal nuclear
matter density (referred to as $\rho _0$), the bag constant, the
total energy of a bag and the quark condensate vanish simultaneously,
and the radius of a nucleon becomes infinite. Such behaviors indicate
that nucleons can no longer exist as bags consisting of quarks.
Since the density $12\rho_0$ is larger than the maximal average hadron
density of neutron stars, which is commonly believed to be about
$10\rho_0$, the presently obtained results show that the nuclear matter
changes to quark matter, i.e., the phase transition from hadrons to
quarks happens, as the density of nuclear matter gets beyond the
maximal average density of neutron stars. It provides a clue that
there may be quark matter or hybrid matter with hadrons and quarks
in the center part of neutron stars. This is something similar to the
result given in Ref.\cite{Pra97}. On the other hand, the quark condensate
$<\overline{q}q>$ has been regarded to be a manifestation to identify
the chiral symmetry breaking. The gradual decrease of the quark
condensate indicates that the chiral symmetry is restored gradually
as the nuclear matter density increases. When the nuclear matter
becomes quark matter, the chiral symmetry is restored completely since
the quark condensate vanishes. It indicates that the quark decoupling
phase transition and the chiral phase transition may happen at the
same nuclear matter density. It is also worth mentioning that the
increase of the nucleon radius induces naturally the swell of nucleons,
which is believed to be essential to the EMC effect (see for example
Ref.\cite{LSYS94}).

To show the function of the bag in nuclear matter and the process of
the phase transition, we evaluate the critical radius of the bag in
nuclear matter and the variation feature of the pressure of the bag
against the nuclear matter density. In a classical point of view,
the relation between the critical radius of the bag $r_c$ and the
nuclear matter density $\rho$ reads
$$ \rho \frac{4}{3} r_c ^3 =1 \, .$$
From the thermodynamics of quark system\cite{Kap89}, the pressure in
a bag with quarks
$$p = \int_0^{\mu} n d \mu - {\cal{B}} \, . $$
The numerical results of the critical radius and the pressure of
the bag are displayed in Figs.5 and 6, respectively. It is evident
that, as the nuclear matter is dilute, the pressure of the bag is
negative so that the bag exists definitely and three quarks are
bound in a bag to form a nucleon. Furthermore, the bags exist
separately from each other, since the radius of a bag is smaller
than the critical radius $r_c$. When the density of nuclear matter
increases, the bags get close to each other. As the nuclear matter
density is about 2 times the normal density of nuclear matter, the
nucleons begin to overlap with each other. However, every nucleon
prefers to exist as an independent bag since the pressure of the
bag is still negative. As the density $\rho > 3 \rho _0$, the
pressure changes to positive. It is definite that such a positive
pressure enhances the overlapping among nucleons. Then quark
matter appears as all the nucleons overlap with each other
completely as the nuclear matter density $\rho > 12 \rho _0$. As a
consequence, the phase transition from hadrons to quarks takes
place. On the other hand, taking the same way as that in
determining the relation between nuclear matter density and quark
chemical potential, and considering the fact that the degeneracy
of nucleons is $g=4$, one can easily get that $ \rho _{_B} =
\frac{2} {3 \pi ^2} \mu _{_B}$. Comparing this result with
Eq.~(23), one can know that the chemical potential of quarks is
just the same as that of nucleons. Such a equivalence means that
the phase transition can happen.

\begin{figure}
\begin{center}
\resizebox{10cm}{!}{\includegraphics{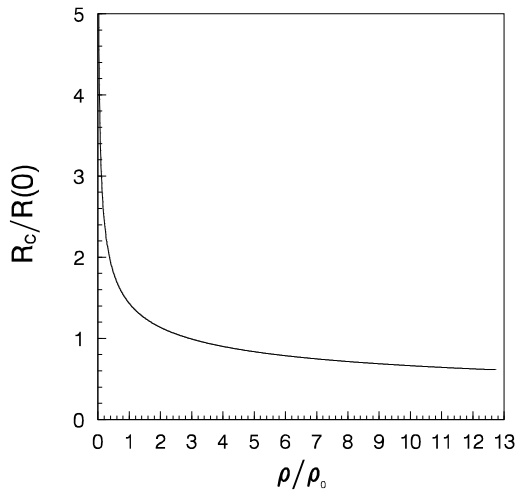}}
\caption{Calculated critical radius of a bag in nuclear matter
with respect to the nuclear matter density} \label{fig:Fig.5}
\end{center}
\end{figure}

\begin{figure}[h]
\begin{center}
\resizebox{10cm}{!}{\includegraphics{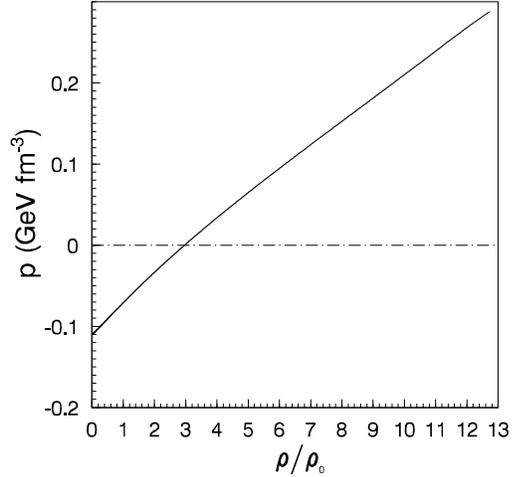}} \caption{
Calculated pressure of the bag with respect to the nuclear matter
density } \label{fig:Fig.6}
\end{center}
\end{figure}

Figures 1 and 2 show also that, before the phase transition takes
place, the changing characteristics of the bag constant and the
radius of a nucleon coincide with the results obtained in the QMC
model with a phenomenological dependence on the scalar meson field
being involved\cite{LTh98,JJ967,MJ978,Guo99}. Such a behavior
indicates that, dealing the bag constant and the bag radius with
a phenomenological dependence on the medium density in the QMC
model is reasonable. Nevertheless, Fig.~3 shows that the change
of the total energy of a bag is much less rapid than that of nucleon
mass obtained in QMC and other frameworks. Tracking down the source
of the discrepancy, one knows that the total energy of a bag is not
identical to the mass of a nucleon, but with a relation $M=\sqrt{E^2
- <p>^2 }$. Taking $<p>^2 = \mu _{_B} ^2$ approximately, we get that,
at the normal density of nuclear matter, the mass of a nucleon
decreases to about 85\% of the value in free space. This is consistent
with the result determined self-consistently in the derivative scaler
coupling model of the RMF\cite{Guo99}. It means then the effective mass
of nucleons in RMF calculations is also reasonable. Such agreements
show that present model can reproduce the quantities and their variation
characteristics obtained in the QMC model and the RMF qualitatively.

\section{Summary and Remarks}

In summary we have investigated the density dependence of the bag
constant of nucleons, the nucleon radius and the total energy of the
bag as well as the quark condensate in nuclear matter in the global
color symmetry model, an effective field theory model of QCD.
A maximal density of nuclear matter, which is larger than 12 times
the normal nucleon density, for the existence of the bag of quarks
is obtained. The calculated results indicate that the bag constant,
the total energy of the bag and the quark condensate decreases
with the increasing of the nuclear matter density before the maximal
density is reached. Meanwhile the size of nucleons swells.
As the maximal density is reached, a phase transition from nucleons to
quark-gluons takes place and the chiral symmetry is restored.
Furthermore, the presently obtained changing features agree with those
obtained in QMC and RMF quite well. In this sense, it provides a
clue of the QCD foundation to the QMC and RMF with a simple effective
field model of QCD.

In the present calculation, the $g^2D(p-q)$ is taken to be proportional
to a $\delta-$function. However, the detailed effects of the running
coupling constant, the gluon propagator $D(p-q)$ and the other degrees
of freedom on the changing feature have not yet been included.
Especially, since the meson fields were taken to be $\sigma = 0$ and
$\pi = 0$ in the bag with respect to those of the vacuum configuration
$\sigma =1$, $\pi =0$, the self-consistent interaction and adjustment
between the quarks and the meson fields have not yet been taken into
account. Meanwhile, the relation between the chemical potential and
the nuclear matter density was handled with a simple correspondence
in statistical mechanics. It means that the present calculation is
an approximation of the scheme. The obtained results are thus the
preliminary ones. In principle, the quark energy (and field), the
meson fields and the nuclear matter density can be determined by
solving the set of differential-integral equations established from
the action(Eq.(1)) of the system consistently. Such a sophisticated
investigation is under progress.

\bigskip

This work is supported by the National Natural Science Foundation of
China and the Foundation for University Key Teacher by the Ministry
of Education. One of the authors (Liu) thanks also the support by the
Funds of the Key Laboratory of Heavy Ion Physics at Peking University,
Ministry of Education, China.
Another author (Guo) thanks also the support by the Major State Basic
Research Development Program under contract No.~G2000077400.
Helpful discussions with Professors F. Wang, X. F. L\"{u}, E. G. Zhao
and P. F. Zhuang are acknowledged with thanks.

\end{document}